\documentclass[sigplan,screen]{acmart}
\usepackage{listings}
\usepackage{hyperref}
\usepackage{graphicx}
\usepackage{enumitem}

\title{E-Path: Equality Saturation for Control-Flow Graphs}
\author{Guillermo Garcia}
\date{May 26, 2026}

\begin{document}

\begin{abstract}

Modern equality saturation systems excel at expression-level rewrites by exploring large spaces of equivalent programs without suffering from the phase-ordering problem. However, these systems struggle to represent equivalence directly over arbitrary control-flow graphs, often requiring normalization into structured or tree-like forms before optimization can occur.

We present the E-Path data structure, a prototype framework for equality-saturation-style optimization over control-flow graphs. Instead of representing congruence between individual expressions, the E-Path records equivalence between instruction sequences embedded within a compiler intermediate representation. In this prototype, E-Path is instantiated over a restricted ANF-based control-flow graph used in a compiler backend, but the model itself is intended to be IR-agnostic.

By treating instruction sequences as the fundamental unit of congruence, the E-Path enables non-destructive optimization of loops and other control-flow structures while preserving multiple equivalent program variants simultaneously. This allows classical CFG optimizations to be expressed as rewrite-driven transformations without destructive mutation of the underlying graph.

\end{abstract}

\maketitle

\section{Introduction}

The Control Flow Graph (CFG), typically used in conjunction with the Static Single Assignment (SSA) form~\cite{cytron1991ssa}, is the standard intermediate representation in optimizing compilers. SSA decouples variable usage from definitions and simplifies data dependency tracking, enabling many classical optimizations. However, most CFG-based optimizations operate destructively, mutating the intermediate representation in place and discarding prior program states.

As a result, optimization quality depends heavily on pass ordering. Different optimization sequences can produce substantially different generated programs, leading to the well-known phase-ordering problem. In practice, compiler pipelines rely on heuristics and fixed pass schedules rather than exploring the space of equivalent transformations.

Equality saturation~\cite{tate2009eqsat} addresses this issue by representing multiple equivalent programs simultaneously within a shared structure and applying rewrites without committing to a single execution path. Existing systems such as E-Graphs~\cite{willsey2021egg} operate primarily on expression trees or DAGs. Although effective for algebraic transformations, they do not naturally extend to general control-flow graphs without prior normalization.

Alternative representations such as the Regionalized Value State Dependence Graph (RVSDG)~\cite{reissmann2020rvsdg} reduce the dependency on the explicit control-flow structure by encoding programs using nested regions and explicit dependencies. However, these representations still require that arbitrary control flow be normalized before optimization.

These limitations suggest the need for a representation that supports non-destructive reasoning directly over control-flow graphs. This motivates the E-Path data structure.

\section{Running Example}

Consider the following loop:

\begin{verbatim}
loop_header(i):
    c = iconst 42
    one = iconst 1
    next_i = add i, one
    loop_back(next_i)
\end{verbatim}

The instruction \texttt{iconst 42} is loop invariant because it does not depend on loop-carried state. A classical Loop-Invariant Code Motion (LICM) pass would rewrite the loop as:

\begin{verbatim}
preheader:
    c = iconst 42

loop_header(i):
    one = iconst 1
    next_i = add i, one
    loop_back(next_i)
\end{verbatim}

Traditional CFG optimizers perform this transformation destructively, overwriting the original representation. In contrast, the E-Path preserves both versions.

\[
P_1 \in P \quad \text{where } P_1 \text{ is obtained from } P_0 \text{ via LICM}
\]

Both \(P_0\) and \(P_1\) remain valid members of the equivalence set, allowing later extraction to choose between them.

\section{Approach}

The current implementation instantiates E-Path over a restricted A-Normal Form (ANF) control-flow graph used in the Crabstar compiler backend. In this IR, each basic block consists of a single instruction followed by a parameterized control-flow terminator. This constraint is an implementation detail of the backend IR and is used to simplify structural matching and rewrite construction; it is not a requirement of the E-Path model itself.

Instead of maintaining expression-level equivalence classes via union-find (e-classes), the E-Path tracks congruence at the level of instruction sequences. The fundamental unit of equivalence is the E-Sequence, a CFG-derived control-flow region represented as a linear sequence of basic blocks.

Although represented linearly, E-Sequences correspond to structured regions of the CFG rather than strictly linear execution traces. Conditional control flow is represented through parameterized branch terminators that reference successor regions in the underlying CFG, while merge blocks define sequence boundaries. As a result, an E-Sequence may represent higher-level control-flow structure without requiring explicit enumeration of every branch-local block within the sequence itself.

A key design choice is that E-Path is monotonic. Existing sequences are never modified. Instead, every rewrite introduces a new E-Sequence into the equivalence set. The result is a persistent search space over equivalent control-flow transformations.

\subsection{Loop-Invariant Code Motion}

Loop-Invariant Code Motion is implemented as a monotonic rewrite pipeline over E-Sequences:

\begin{enumerate}[label=\arabic*.]
    \item \textbf{Cycle Detection:} The compiler identifies cyclic structure in the E-Sequence corresponding to a loop region in the CFG.
    \item \textbf{Invariance Verification:} A block is considered loop-invariant if its operands and side effects do not depend on values modified within the cycle, according to the IR’s dependency analysis.
    \item \textbf{Sequence Reconstruction:} A new E-Sequence is constructed where invariant blocks are placed in a preheader region before the loop header.
\end{enumerate}

This produces a new equivalent sequence while preserving the original. The result is a growing set of structurally distinct but semantically related control-flow variants.

\section{Formal Model}

\subsection{Control Flow Graph}

A control-flow graph is defined as:

\[
G = (V, E)
\]

where \(V\) is a set of basic blocks and \(E\) is a set of directed edges representing control transfers.

Each block \(b \in V\) contains a single instruction and a terminating control-flow operation.

\subsection{E-Sequences}

An E-Sequence is defined as:

\[
S = [b_1, b_2, \dots, b_n]
\]

where each \(b_i \in V\).

E-Sequences represent CFG-derived control-flow segments that may correspond to paths or structured regions, depending on the shape of the underlying control-flow. While linear, they implicitly encode branching structure through terminator semantics.

\subsection{Semantic Equivalence}

Equivalence is not derived internally by the E-Path. Instead, equivalence is introduced exclusively through externally verified rewrite rules over the intermediate representation.

This places E-Path in the same category as equality saturation systems based on E-Graphs: correctness depends on the soundness of rewrite rules rather than internal proof construction.

\subsection{E-Path Structure}

The E-Path is defined as a monotonic set:

\[
P = \{S_1, S_2, \dots, S_n\}
\]

Each rewrite rule defines a transformation:

\[
S_i \xrightarrow{r} S_j
\]

where \(r\) is a validated control-flow rewrite. All such results are inserted into the same equivalence set.

\section{Rewrite System}

Because the system is monotonic, optimization is formulated as repeated application of rewrite rules that expand the set of known equivalent E-Sequences. Non-destructive allows for optimization opportunities to be introduced by one transformation being unable to invalidate or erase previously discovered variants. This allows extraction to reason globally across multiple competing control-flow organizations rather than committing to a single optimization trajectory early in compilation.

\subsection{LICM as a Rewrite Rule}

Loop-Invariant Code Motion is expressed as a transformation over cyclic E-Sequences. The rule can be informally written as:

\[
\text{loop}(I, B_{inv} \cup B_{var}) \rightarrow B_{inv}; \text{loop}(I, B_{var})
\]

where \(B_{inv}\) are invariant blocks and \(B_{var}\) are loop-dependent blocks.

This transformation does not replace the original sequence but adds a structurally distinct equivalent sequence.

\section{Cost Extraction}

Because multiple equivalent E-Sequences coexist, extraction selects a final program using symbolic cost evaluation.

Loop cost is modeled as:

\[
C = N \cdot M
\]

where \(N\) is a symbolic iteration count and \(M\) is the aggregate cost of a loop body.

Costs compose over sequences by summing block costs and scaling loop regions. This yields a comparison function over entire control-flow variants rather than local instruction costs.

Extraction evaluates all candidate E-Sequences and selects:

\[
S^* = \arg\min_{S \in P} C(S)
\]

\section{Pattern Matching}

The system supports two complementary matching modes.

\subsection{Expression-Level Matching}

Because the IR is in ANF, data dependencies are explicit. This allows instruction sequences to be matched structurally as computation graphs, similar to expression matching in traditional E-Graphs.

\subsection{Control-Flow Matching}

In parallel, pattern matching operates over CFG topology. The current implementation supports:

\begin{itemize}
    \item Acyclic instruction sequences
    \item Reducible loop regions
\end{itemize}

These structures form the current basis of the rewrite and matching system.

\section{Architectural Trade-offs}

The E-Path is best understood as a persistent search space over CFG variants rather than a minimal representation. The monotonic nature of E-Path implies unbounded growth in the number of stored E-Sequences. To address this, the implementation uses hash consing and structural deduplication. 

Duplicate sequences are rejected based on structural hashing, ensuring the equivalence set remains finite for a given rewrite closure. 

Termination is defined as reaching a fixed point where no new E-Sequences are generated.

A prototype implementation exists in Rust as part of the Crabstar compiler backend.

\section{Related Work}

Equality saturation systems such as E-Graphs~\cite{tate2009eqsat,willsey2021egg} represent equivalence over expressions using shared equivalence classes. These systems are effective for algebraic transformations but do not naturally extend to general control-flow structures.

RVSDGs~\cite{reissmann2020rvsdg} eliminate explicit CFG structure by encoding programs as nested regions, but still require normalization of arbitrary control flow before optimization.

Traditional SSA-based compilers operate directly on CFGs but rely on destructive transformation pipelines, making optimization sensitive to pass ordering.

E-Path differs by maintaining equivalence directly over executable instruction sequences embedded in a CFG structure, enabling non-destructive exploration of control-flow variants. 

\section{Current Limitations}

The current implementation supports only reducible control-flow structures and does not yet model aliasing, memory effects, or speculative execution.

Additionally, the rewrite system assumes correctness of external equivalence proofs and does not internally verify semantic preservation.

\section{Conclusion}

The E-Path data structure extends equality saturation to control-flow graphs by treating instruction sequences as the unit of congruence. This enables non-destructive optimization over CFG structure without requiring normalization into tree-based representations.

By preserving multiple equivalent control-flow variants simultaneously, E-Path reframes classical compiler optimizations such as LICM as monotonic rewrites over a shared search space. This suggests that phase-ordering problems in CFG optimization can be treated as global search rather than sequential transformation.

\section{Future Work}

The current system allows for recognition of acyclic sequences of instructions and reducible loops. Future work will incorporate more complex control flow structures into the pattern matching engine such as conditional branching, jump tables and irreducible loops. This would allow for more complex optimizations such as branch distribution, loop unswitching and dead code elimination.

Because E-Sequences represent control flow segments, they can be optimized in ways similar to traditional destructive rewrites while maintaining the monotonic properties of an E-Graph. Future work will introduce aggressive optimizations, especially in regard to loop optimizations. Loop fusion, loop fission, loop interchange, partial unrolling and vectorization are all planned to be implemented as rewrite rules. In addition, other algebraic optimizations such as constant folding, constant propagation, dead code elimination, and common subexpression elimination are also planned.

Due to the monotonic nature of the E-Path, it could be better suited for parallelism. Because each E-Sequence being an independent candidate for rewrite application, pattern matching and rewrites could be run in parallel, with the synchronization point being insertion into the E-Path’s Hash Map for equivalencies. Similarly, the cost extraction could also be written to be parallel because of the same properties, allowing each equivalent sequence’s cost to be calculated independently. This design supports a parallel search strategy in which multiple control-flow variants are explored simultaneously, potentially scaling to large optimization spaces without centralized bottlenecks.

\bibliographystyle{plain}
\bibliography{references}
\end{document}